
\input epsf

\ifx\epsffile\undefined\message{(FIGURES WILL BE IGNORED)}
\def\insertfig#1#2{}
\def\insertfigsmall#1#2{}
\else\message{(FIGURES WILL BE INCLUDED)}
\def\insertfig#1#2{{{\baselineskip=4pt
\midinsert\centerline{\epsfxsize=\hsize\epsffile{#2}}{{
\centerline{#1}}}\medskip\endinsert}}}
\def\insertfigsmall#1#2{{{\baselineskip=4pt
\midinsert\centerline{\epsfxsize=4.5in\epsffile{#2}}{{
\centerline{#1}}}\medskip\endinsert}}}
\def\insertfigtiny#1#2{{{\baselineskip=4pt
\midinsert\centerline{\epsfxsize=2.8in\epsffile{#2}}{{
\centerline{#1}}}\medskip\endinsert}}}
\fi

\input harvmac

%
%
\ifx\answ\bigans
\else
\output={
  \almostshipout{\leftline{\vbox{\pagebody\makefootline}}}\advancepageno
}
\fi
%
%
%

%
%

%
%
\def\UCSD#1#2{\noindent#1\hfill #2%
\bigskip\supereject\global\hsize=\hsbody%
\footline={\hss\tenrm\folio\hss}}
%
%
\def\abstract#1{\centerline{\bf Abstract}\nobreak\medskip\nobreak\par #1}
%
%
%
%
\edef\tfontsize{ scaled\magstep3}
 \tfontsize  \tfontsize
 \tfontsize \font\titlei=cmmi10 \tfontsize
\font\titleis=cmmi7 \tfontsize \font\titleiss=cmmi5 \tfontsize
\font\titlesy=cmsy10 \tfontsize \font\titlesys=cmsy7 \tfontsize
\font\titlesyss=cmsy5 \tfontsize  \tfontsize
\skewchar\titlei='177 \skewchar\titleis='177 \skewchar\titleiss='177
\skewchar\titlesy='60 \skewchar\titlesys='60 \skewchar\titlesyss='60
%
%
%
%
%
\def\inv{^{\raise.15ex\hbox{${\scriptscriptstyle -}$}\kern-.05em 1}}
\def\lbar{{\lower.35ex\hbox{$\mathchar'26$}\mkern-10mu\lambda}}

%
%
%
%
\def\dsl{\,\raise.15ex\hbox{/}\mkern-13.5mu D}
\def\delsl{\raise.15ex\hbox{/}\kern-.57em\partial}
\def\Ksl{\hbox{/\kern-.6000em\rm K}}
\def\Asl{\hbox{/\kern-.6500em \rm A}}
\def\Dsl{\hbox{/\kern-.6000em\rm D}} 
\def\Qsl{\hbox{/\kern-.6000em\rm Q}}
\def\gradsl{\hbox{/\kern-.6500em$\nabla$}}
%
%
\def\lspace{\ifx\answ\bigans{}\else\qquad\fi}
\def\lbspace{\ifx\answ\bigans{}\else\hskip-.2in\fi} 
%
%
\def\boxeqn#1{\vcenter{\vbox{\hrule\hbox{\vrule\kern3pt\vbox{\kern3pt
        \hbox{${\displaystyle #1}$}\kern3pt}\kern3pt\vrule}\hrule}}}
%
%
\def\mbox#1#2{\vcenter{\hrule \hbox{\vrule height#2in
\kern#1in \vrule} \hrule}}
%
%
%
%

%
%
%
%
%

%

%
%

\def\darr#1{\raise1.5ex\hbox{$\leftrightarrow$}\mkern-16.5mu #1}

%
%
\def\frac#1#2{{\textstyle{#1\over #2}}} 
%
%
%
%

%
%
%
%

%
%
\def\ltap{\ \raise.3ex\hbox{$<$\kern-.75em\lower1ex\hbox{$\sim$}}\ }
\def\gtap{\ \raise.3ex\hbox{$>$\kern-.75em\lower1ex\hbox{$\sim$}}\ }
\def\gl{\ \raise.5ex\hbox{$>$}\kern-.8em\lower.5ex\hbox{$<$}\ }
\def\roughly#1{\raise.3ex\hbox{$#1$\kern-.75em\lower1ex\hbox{$\sim$}}}
%
%

%

%
\def\np#1#2#3{{Nucl. Phys. } B{#1} (#2) #3}
\def\pl#1#2#3{{Phys. Lett. } {#1}B (#2) #3}

\def\physrev#1#2#3{{Phys. Rev. } {#1} (#2) #3}

\relax

\def\six{${\bf 6}$}

\def\three{${\overline{\bf 3}}$}
\def\lamc{$\Lambda_c^+$}
\def\xictp{$\Xi_{c3}^+$}
\def\xictz{$\Xi_{c3}^0$}
\def\sigcpp{$\Sigma_c^{++}$}
\def\sigcp{$\Sigma_c^+$}
\def\sigcz{$\Sigma_c^0$}
\def\xicsz{$\Xi_{c6}^0$}
\def\xicsp{$\Xi_{c6}^+$}
\def\omc{$\Omega_c^0$}
\def\ssg{$\Sigma_Q^{*}\rightarrow\Lambda_Q\gamma$}
\def\ssc{$\Sigma_c^{*}\rightarrow\Lambda_c\gamma$}
\def\sigc{$\Sigma_c$}
\def\sigcs{$\Sigma_c^*$}

\def\vslash{v\hskip-0.5em /}
\hbadness=10000

\noblackbox
\vskip 1.in
\centerline{{\titlefont{ E2 Strength in the Radiative  }}}
\medskip
\centerline{{\titlefont{Charmed Baryon Decay \ssc\  }
\footnote{$^\dagger$}{{\tenrm Work
supported in part by the Department of Energy under contract
DE--FG02--91ER40682.}}}}
\vskip .5in
\centerline{{Martin J.~Savage}}
\medskip
{\it{\centerline{ Department of Physics, Carnegie Mellon University,
Pittsburgh PA 15213}}}

\vskip .2in

\abstract{
The radiative decay \ssg\ can have both magnetic dipole (M1) and electric
quadrupole (E2) components.  In the heavy quark limit $M_Q\rightarrow\infty$
the transition arises from the spin of the light degrees of freedom changing
from $s_l=1$ to $s_l=0$ and hence the E2 contribution vanishes.
We compute the leading contribution to the E2 strength in chiral
perturbation theory and find that the amplitude is enhanced by a small energy
denominator in the chiral limit. This enhancement essentially compensates
for the  $1/M_c$ suppression that  is present in the charm system.
We find a mixing ratio of order a few percent dependent upon the
\sigc-\sigcs\ spin symmetry breaking mass difference.
The analogous quantity in the b-baryon sector is smaller by a factor of
$\sim M_c/M_b$.
 }

\vfill
\UCSD{\vbox{
\hbox{CMU-HEP 94-25}
\hbox{DOE-ER/40682-79}
\hbox{hep-ph/9408294}
}}{August 1994}
\eject

It has recently been shown that the leading contribution to the electric
quadrupole (E2) strength of a radiative transition between light baryons,
such as $\Delta\rightarrow N\gamma$, can be computed in chiral perturbation
theory
\ref\bss{M.N. Butler, M.J. Savage and R.P. Springer, \pl{304}{1993}{353};
\pl{314}{1993}{122} (E+A).}.
It was found that such matrix elements are enhanced in the chiral limit  by
factors of  ${\sim\log\left( m_\pi^2/\Lambda_\chi^2 \right) }$ (where
$\Lambda_\chi$ is the chiral symmetry breaking scale) giving mixing
ratios (ratio of E2 to M1 matrix elements) typically of order a few percent.

The situation is somewhat more complicated for transitions between baryons
containing a single heavy quark.   In the limit that the heavy quark mass
is infinitely greater than the scale of strong interactions
($M_Q\gg\Lambda_{\rm
QCD}$) the lowest lying baryons containing one heavy quark can be
classified by the spin of the light degrees of freedom.  In the charm systems,
the \three\  of charmed baryons (\lamc\ ,\ \xictp\ and
\xictz\ ) have the light degrees of freedom in a spin zero configuration
($s_l=0$) while the \six$^{(*)}$\  of charmed baryons (\sigcpp$^{(*)}$\ ,\
\sigcp$^{(*)}$\ ,\
\sigcz$^{(*)}$\ ,\ \xicsp$^{(*)}$\ ,\ \xicsz$^{(*)}$\ and \omc$^{(*)}$\ )
have the light degrees of freedom in  a spin one configuration ($s_l=1$).

The radiative decay \ssg\ ($Q=c, b$) involves a $J={3\over 2}$ state with
$s_l=1$
decaying into a $J={1\over 2}$ state with $s_l=0$.  The heavy quark is  a
spectator
during the transition and the decay proceeds entirely through the change of
spin
of the light degrees of freedom, allowing only magnetic dipole (M1) radiation.
It is clear that any E2 radiation can only arise from finite mass effects
of the heavy quark.  Naively this suggests that the E2/M1 mixing ratio $\delta$
will
be much smaller in heavy baryons than the corresponding quantity in the light
baryon
sector. However, we will show that in fact the E2 matrix element is enhanced by
a
small energy denominator in the chiral limit that essentially compensates for
the
$1/M_c$ suppression in the charmed baryon sector.
A thorough investigation of the M1 radiative decays of charmed baryons
can be found in
\ref\ccllyy{H.Y. Cheng etal, \physrev{47}{1993}{1030}.}
\ref\chogeorgi{P. Cho and H. Georgi,
\pl{296}{1992}{408};\pl{300}{1993}{410}(E).}
but there has been no discussion of the contribution from higher
multipoles where appropriate. With the fixed target experiment E781
scheduled to run in 1996 that is capable of measuring the angular
distribution of the
photons in charmed baryon decays it seems timely to try and better
understand such processes.   In this work we will compute the leading
contribution to the E2 matrix elements for \ssg\ in chiral perturbation theory.

The strong interaction dynamics of quarks greatly simplifies in the limit
that their mass becomes infinitely greater than the scale of strong
interactions
\ref\iw{N. Isgur and M.B. Wise, \pl{232}{1989}{113};
\pl{237}{1990}{527}.}\nref\eh{E. Eichten and B. Hill,
\pl{234}{1990}{511}.}-\ref\hg{H. Georgi, \pl{240}{1990}{447}.}.  The new
symmetries that become manifest in this limit have been been combined with
chiral
symmetry order to describe the soft hadronic interactions of hadrons containing
a
single heavy quark
\ref\mbw{M.B. Wise, \physrev{45}{1992}{2188}.}\nref\bd{G. Burdman and J.
Donohue, \pl{208}{1992}{287}.}\nref\yan{T.M. Yan etal,
\physrev{46}{1992}{145}.}-\ref\cho{P. Cho, \pl{285}{1992}{145}.}
(for a review see \ref\mbwrev{M.B. Wise, Lectures given at the CCAST
Symposium on Particle Physics at the Fermi Scale, May 1993.} ).
The lowest lying \three\ of  baryons containing a heavy quark
is described by the field $T_i(v)$ where $v$ is the four-velocity of the baryon
(conserved during soft hadronic interactions) and where $i$ is the SU(3)
index.
Similarly, the lowest lying \six$^{(*)}$\ of baryons containing a heavy quark
is
denoted by the field $S(v)^{ij}_\mu$ (symmetric on $i, j$) where $i,j$ are
SU(3)
indices and where $\mu$ is a lorentz index.
The chiral lagrangian describing the lowest order soft hadronic
interactions of these baryons is given by (using the notation of \cho\ )
\eqn\soft{ \eqalign{ {\cal L} = & \overline{T}^i iv\cdot D T_i
- \overline{S}^\mu_{ij} iv\cdot D S_\mu^{ij}
+\Delta_0 \overline{S}^\mu_{ij} S_\mu^{ij}  \cr
& + g_3\left( \epsilon_{ijk} \overline{T}^i ( A^\mu )^j_l S^{kl}_\mu
+ {\rm h.c.} \right) +
ig_2\epsilon_{\mu\nu\rho\sigma}\overline{S}^\mu_{ij}v^\nu ( A^\rho )^i_k
S^{\sigma, jk}}\ \ \ \ .}
The axial field of pseudo-Goldstone bosons
$A_\mu = {i\over 2}\left(\xi\partial_\mu\xi^\dagger -
\xi^\dagger\partial_\mu\xi\right)$ is defined in terms of
$\xi=\exp\left( iM/f\right)$ where $M$ is the octet of pseudo-Goldstone
bosons represented by
\eqn\mesons{M=\left(\matrix{
\eta/\sqrt{6}+\pi^0/\sqrt{2} & \pi^+ & K^+ \cr
\pi^- & \eta/\sqrt{6}-\pi^0/\sqrt{2} & K^0 \cr
K^- & \overline{K}^0 & -2\eta/\sqrt{6} } \right)  \ \ \ .}
The residual mass $\Delta_0$ (mass difference between the \three\ and
\six\ ) is present even in the infinite mass limit and arises from the
energy difference between the light degrees of freedom in the
\three\ and \six\ .
Using the notation of \cho\ we write
\eqn\fields{\eqalign{  T_i(v) = & {1\over 2} (1+\vslash ) B_i \cr
S_\mu^{ij}(v) = & {1\over\sqrt{3}}(\gamma_\mu+v_\mu)\gamma_5 {1\over 2}
(1+\vslash ) B^{ij} + {1\over 2} (1+\vslash ) B^{*ij}_\mu }\ \ \ .}
The charmed baryons in the \three\  have SU(3) assignments
\eqn\cthree{ B_1=\Xi_{c3}^0\ ,\ \ B_2=-\Xi_{c3}^+\ ,\ \
B_3=\Lambda_c^+\ \ \ ,} while the $J=1/2$ charmed baryons in the \six\
have SU(3) assignments
\eqn\csix{\eqalign{
B^{11} = \Sigma_c^{++}\ ,\ \  B^{12}={1\over\sqrt{2}}\Sigma_c^+\ ,\ \
B^{22}=\Sigma_c^0  \cr
B^{13}={1\over\sqrt{2}}\Xi_{c6}^+\ ,\ \  B^{23}={1\over\sqrt{2}}\Xi_{c6}^0\ ,\
\
B^{33}=\Omega_c^0 }\ \ \ .}
The $J=3/2$ members of the \six$^*$\ have the same SU(3) assignments in
$B^{*ij}_\mu$ as their $J=1/2$ partners have in $B^{ij}$.
We will need the strong decay widths of the $\Sigma_c^{*+}$
later in this work and they can be shown to be
\eqn\stronga{ \Gamma ( \Sigma_c^+ \rightarrow\Lambda_c\pi^0 )  =
\Gamma ( \Sigma_c^{*+} \rightarrow\Lambda_c\pi^0 ) =
{g_3^2\over 6 \pi f_\pi^2 } |{\bf k}_\pi|^3 \ \ \ ,}
and
\eqn\strongb{ \Gamma ( \Sigma_c^{*+} \rightarrow\Sigma_c\pi ) =
{g_2^2\over 18 \pi f_\pi^2 } |{\bf k}_\pi|^3 \ \ \ ,}
where ${\bf k}_\pi$ is the pion three-momentum.

A complete discussion of the M1 decays of baryons containing a single heavy
quark can
be found in \ccllyy\chogeorgi\  .  The width for the decay \ssg\ is given by
\eqn\width{\Gamma (\Sigma_Q^{*}\rightarrow\Lambda_Q\gamma) =
{E_\gamma^3\over 12\pi} \left[ |A_1|^2 + 3|A_2|^2 \right]\ \ \ ,}
where $A_1$ is the M1 amplitude, $A_2$ is the E2 amplitude and $E_\gamma$
is the energy of the emitted photon.  The E2/M1 mixing ratio is defined by
\eqn\mix{\delta = {A_2\over A_1}\ \ \ .}

In chiral perturbation theory the M1 matrix element is dominated by a dimension
five local counter term,
\eqn\mop{ {\cal O}_{\rm M1} \sim {e\over \Lambda_\chi}
\overline{\Lambda}_Q {\cal Q} \gamma^\mu\gamma_5\Sigma_Q^{* \nu} F_{\mu\nu} \ \
\ \ ,}
where ${\cal Q}$ is the electromagnetic charge matrix for the three light
quarks given by
\eqn\charge{{\cal Q} = {1\over 3}\left( \matrix{2 & 0 & 0 \cr 0 & -1 & 0\cr 0 &
0
& -1} \right)\ \ \ .}
Loops and higher dimension operators are suppressed by additional powers of
$\Lambda_\chi$.
As this operator has an unknown coefficient and there is no experimental data
on this decay mode we cannot yet determine the M1 matrix element.
However, the M1 matrix element $A_1$ has been computed in the non-relativistic
quark model (NRQM) \ccllyy\  and is found to be
\eqn\nrqm{A_1^{\rm NRQM} = -{e\over M_U}\ \ \ ,}
where $M_U$ is the mass of a constituent up quark $\sim 300$ MeV
(there is no contribution from a heavy quark interaction since the spin of the
light degrees of freedom must change during the transition).
The NRQM estimate should be relatively reliable as it generally
reproduces well the value of M1 matrix elements corresponding to simply
spin-flip
(short  distance)
transitions.  Of course, this is only a guide and ultimately the M1  matrix
elements will be determined directly from experiment. More importantly, the
NRQM
is found to predict very small E2 amplitudes, as they arise from ground state
configuration mixing (see for example
\ref\cap{S. Capstick, \physrev{46}{1992}{1965}.}).
In contrast, the estimates found in chiral perturbation theory are relatively
large \bss\
and result from long-distance charged pion configurations (loops) that are not
present
in the NRQM.

The leading contribution to the E2 matrix element for \ssg\ arises from
the graph shown in
\fig\graph{The graph giving the leading contribution to the E2 matrix
element for \ssg\ . There is a cancellation between the $\Sigma_Q^*$
and $\Sigma_Q$  intermediate states that becomes exact as they become
degenerate.} .
There is a cancellation between the $\Sigma_Q^*$ and $\Sigma_Q$
intermediate states that becomes exact as the states become degenerate.
This is a result of the heavy  quark spin
symmetry which is broken by the $\Sigma_Q^*-\Sigma_Q$ mass splitting,
denoted by $\Delta_Q$.
At this order there is no contribution from $\Lambda_Q$  intermediate states
as its coupling to pions is explicitly suppressed by $1/M_Q$.
We find that the E2 amplitude is given by
\eqn\loopmix{ A_2 = {2\sqrt{2}\over 3} {e g_2 g_3\over 16\pi^2f_\pi^2}
F(\Delta_Q, E_\gamma, m_\pi )\ \ \ ,}
where the function $F(\Delta, E, m )$ is
\eqn\func{\eqalign{
F(\Delta, E, m ) = & \int_0^1\ dx\ (1-2x)\  \left[ J(xE-\Delta,  m ) - J(xE, m
) \right]  \cr
J(y, m ) = & \sqrt{y^2 - m^2}\ \log\left({
y + \sqrt{y^2 - m^2 + i\epsilon } \over y - \sqrt{y^2 - m^2 + i\epsilon}
}\right) }\ \ \ .}
In the limit that $\Delta \ll E_\gamma, m_\pi$ we can expand
the function  and find that
\eqn\expand{F(\Delta, E, m ) = \int_0^1\ dx\ x(1-2x)
{E\Delta\over\sqrt{x^2E^2-m^2}}
\log\left({ x E  + \sqrt{x^2E^2 - m^2 + i\epsilon } \over x E
- \sqrt{x^2E^2 - m^2 + i\epsilon} }\right)\ \ \ .}
which has the limits
\eqn\limits{\eqalign{
F(\Delta, E, m )\ \rightarrow &\  {\pi\over 6}\ {\Delta\over m_\pi}\  E\ \ \
{\rm for }\ \  E\ll m_\pi  \cr
\rightarrow & \  {\Delta\over E}\  E \ \ \ \ \ \ \ {\rm for }\ \  E\gg  m_\pi}\
\ \ \ .}
We see that the infrared divergence of the loop graph is regulated by
the larger of $m_\pi$ and $E_\gamma$, giving a small energy denominator.
We have kept a factor of $E$ in \limits\   because the E2 operator
\eqn\quadrupole{ {\cal O}_{\rm E2} \sim {e\over \Lambda_\chi^2} \
\overline{\Lambda}_Q {\cal Q} \gamma^\mu\gamma_5\Sigma_Q^{* \nu} \
v^\alpha \  \left( \partial_\mu  F_{\alpha\nu} \ +\  \mu\leftrightarrow\nu
\right)
\ \ \ ,}
is dimension six  and when  expanded has an explicit factor of
$E_\gamma /\Lambda_\chi^2$.
Naively one would expect this E2 operator to have a coefficient suppressed by
a factor  of $\Lambda_{\rm QCD}/M_Q$, giving a mixing ratio smaller
than that found  from the long-distance loop graph.

We now turn to the charmed baryon sector  and make an estimate of $\delta$
for \ssc\ .
The $\Sigma_c^*$ has not been observed to date and so the photon energy
for the transition is an unknown, as is the size of the $\Sigma_c^*-\Sigma_c$
spin-symmetry breaking mass difference.
Further, the axial coupling constants $g_{2,3}$ have not been determined
experimentally.
Recently, it has been shown that they are related to the $NN\pi$ coupling
constant,
$g_A=1.25$ in the large-$N_C$ limit of QCD ($N_C$ is the number of colours) and
we will use this estimate of
$g_3=\sqrt{3/2}g_A$ and $g_2=-3/2g_A$
\ref\glm{Z. Guralnik, M. Luke and A.V. Manohar, \np{390}{1993}{474}.}
\ref\jenk{E. Jenkins, \pl{315}{1993}{431}.}
in our computations.
Using the NRQM estimate of the M1 strength and the large-$N_c$
values of $g_{2,3}$ we find that the radiative branching fraction is
 $\sim 1\%$ for $M_{\Sigma_c^*}-M_{\Sigma_c} = 40 {\rm MeV}$
and doesn't vary much over a reasonable range for
$M_{\Sigma_c^*}-M_{\Sigma_c} $.
The E2/M1 mixing ratio $\delta$  is shown in
\fig\mixfig{The E2/M1 mixing ratio $\delta$ for \ssc\ as a function of
$M_{\Sigma_c^*}-M_{\Sigma_c}$ for $g_3=\sqrt{3/2}\ g_A$\ , $g_2=-3/2\ g_A$
and $g_A=1.25$.   An imaginary part (dashed line) arises when
$M_{\Sigma_c^*}-M_{\Sigma_c} > m_\pi$. }
as a function of $M_{\Sigma_c^*}-M_{\Sigma_c}$ for the large-$N_C$
values of $g_2$ and $g_3$.
When $M_{\Sigma_c^*}-M_{\Sigma_c} > m_\pi$ an imaginary part develops due to
the
presence of  $\Sigma_c\pi$ physical intermediate states in the loop graph.
We see from \mixfig\ that  an E2/M1 mixing ratio of order a percent or  so is
expected.

In order to determine $\delta$ in the radiative decay \ssg\  the initial
$\Sigma_Q^*$'s
must be aligned,
by which we mean that the probablility of being in the $m = \pm {3\over 2}$
angular
momentum states $w(\pm {3\over 2})$ are different from the probablility of
being in
the $m = \pm {1\over 2}$ angular momentum states
$w(\pm {1\over 2})$.
The resulting angular distribution of photons from the decay  \ssg\ of an
aligned
$\Sigma_Q^*$ has the form \bss
\ref\rb{H.J. Rose and D.M. Brink, Rev. of Mod. Phys. {\bf 39} (1967) 306.}\
\eqn\dist{ S(\theta) = 1 +
\left[  w(\pm {3\over 2}) - w(\pm {1\over 2}) \right]
{  1 + 6 Re(\delta) - 3|\delta|^2  \over   2 (1 + 3|\delta|^2 )  }
P_2 (\cos\theta)\ \ \ \ ,}
where $P_2 (x)$ is the second Legendre polynomial and $\theta$ is the angle
between the
alignment axis and the photon direction.
At present the alignment of charmed baryons that will be produced in E781
cannot be
predicted and will have to be determined experimentally.
However,  if the polarisations observed in the production of strange baryons
are an indication
of what we may expect in the charm sector then the $\Sigma_c^*$'s could  have
substantial alignment.

It must be remembered that we have computed only the leading contribution
to the mixing ratio.   There will be contributions from other $1/M_c$ operators
that
are formally suppressed in the chiral and heavy quark limits but in reality may
have a
nonnegligible contribution.  We therefore must regard this calculation as
an estimate only and an example of an additional $1/M_c$ contribution is from
loop graphs with  the $\Lambda_c$ as an intermediate state.
Its coupling to pions is  explicitly $1/M_c$ and we expect the loop graph to be
of order
$\sim\log \left( m_\pi^2/\Lambda_\chi^2 \right)$ for $E_\gamma \ll m_\pi$, this
is  less
singular
in the chiral limit  than the loop graph we have computed (this graph does have
an imaginery part for all values of $M_{\Sigma_c^*}-M_{\Sigma_c}$ as
$M_{\Sigma_c^*}-M_{\Lambda_c} > m_\pi$).
However, we know that $E_\gamma\sim m_\pi$ and these additional
contributions may be significant.
Our results, of course, can be applied to the b-baryons by replacing
$M_{\Sigma_c^*}-M_{\Sigma_c}$ with $M_{\Sigma_b^*}-M_{\Sigma_b}$ where
appropriate. Hence, the E2 component is reduced in the b-systems by a
factor of $\sim M_c/M_b$, while the M1 component is essentially unchanged.

In conclusion, we have computed the leading contribution to the E2/M1
mixing ratio for the radiative decays \ssc\ .   We find that the
$1/M_c$ suppression of the E2 amplitude is compensated for by a small energy
denominator arising from the infrared behaviour of pion loop graphs in chiral
perturbation theory.   This indicates that the  E2/M1 mixing ratio could be a
few percent,
depending on the $\Sigma_c^*-\Sigma_c$ spin  symmetry breaking mass splitting.
With fixed target experiments  dedicated to charmed hadrons due to come on line
within the next few years it is possible that this mixing ratio will be
determined and
compared with our estimates.
The corresponding quantity in the b-baryon system will be reduced by a factor
of
$\sim M_c/M_b$.

\bigskip
I would like to thank J. Russ and M. Procario for useful
discussions and bringing my attention to the possibility of measuring the
E2/M1 mixing ratio.  I would also like to thank P. Geiger for useful
discussions.
This work is supported in part by the Department of Energy under
contract DE-FG02-91ER40682 and in part by the NSF under grant  PHY89-04035.

\listrefs
\listfigs
\vfill\eject

\insertfigtiny{Figure 1}{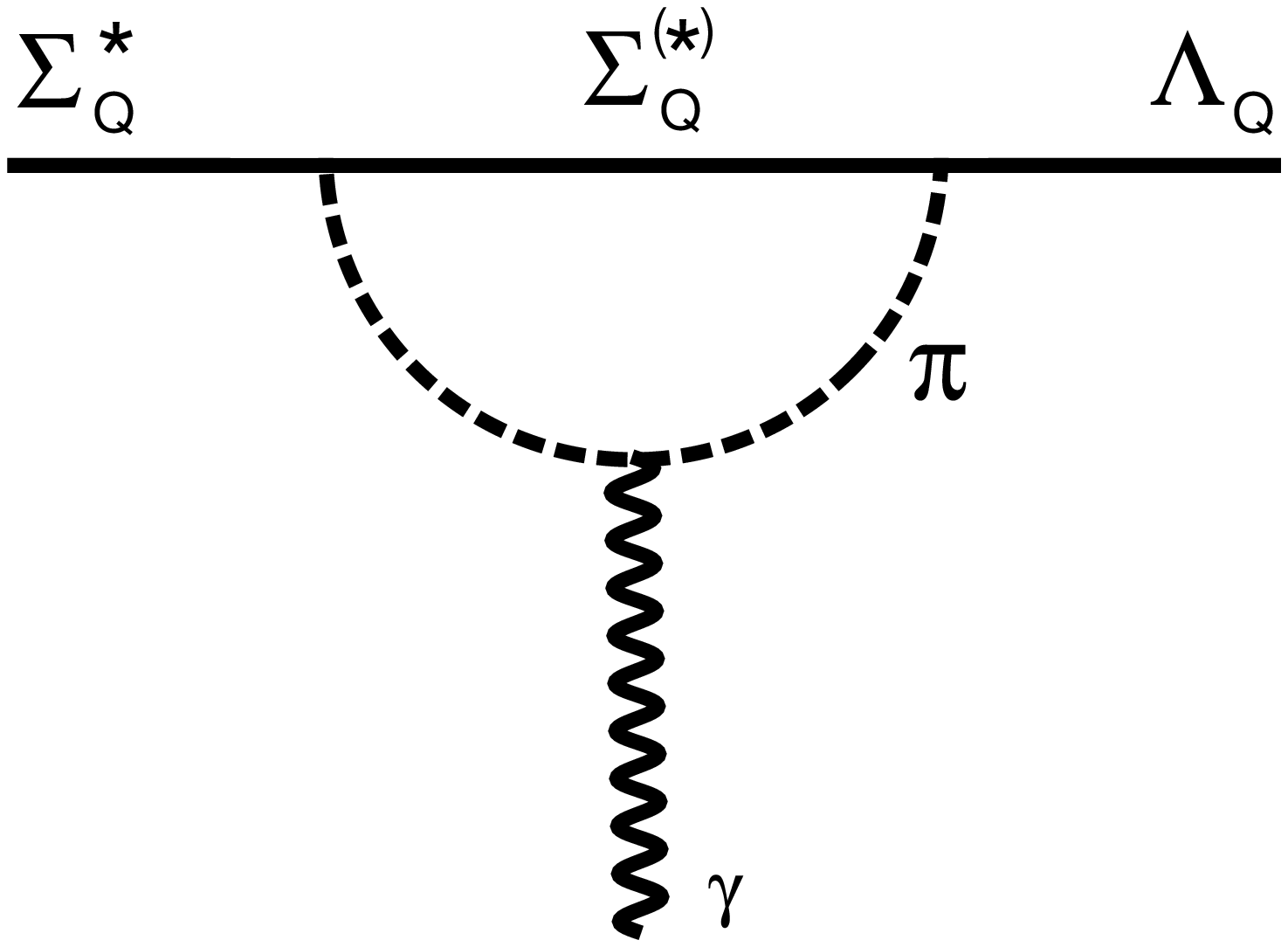}
\vskip 0.5in
\insertfigsmall{Figure 2}{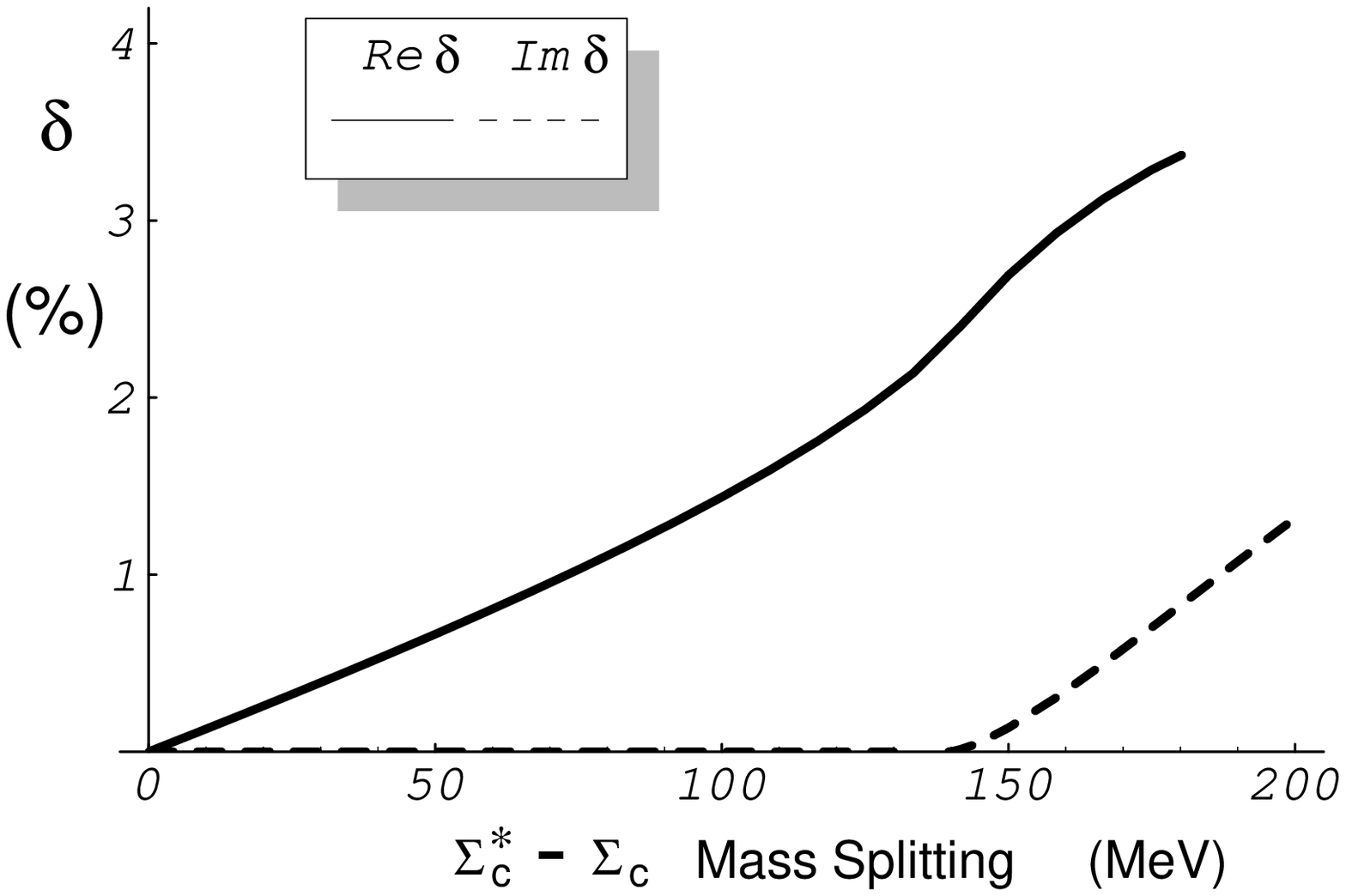}

\bye